\pdfoutput=1
\documentclass[11pt]{article}
\usepackage[final]{acl}
\usepackage{times}
\usepackage{latexsym}
\usepackage{booktabs} 
\usepackage{multirow}
\usepackage[T1]{fontenc}
\usepackage[utf8]{inputenc}
\usepackage{microtype}
\usepackage{inconsolata}
\usepackage{graphicx}

\title{OpenResearcher: Unleashing AI for Accelerated Scientific Research}

\author{Yuxiang Zheng\textsuperscript{\rm{1,8}}\footnotemark[1] \ Shichao Sun\textsuperscript{\rm{4,8}}\footnotemark[1] \ Lin Qiu\textsuperscript{\rm{1}} \ Dongyu Ru\textsuperscript{\rm{1}} \ \textbf{Cheng Jiayang}\textsuperscript{\rm{5}}  \
\textbf{Xuefeng Li}\textsuperscript{\rm{1,8}} \\
\textbf{Jifan Lin}\textsuperscript{\rm{1,8}} \ \textbf{Binjie Wang}\textsuperscript{\rm{3,8}} \ \textbf{Yun Luo}\textsuperscript{\rm{6}}  \ \textbf{Renjie Pan}\textsuperscript{\rm{1}}  \ \textbf{Yang Xu}\textsuperscript{\rm{1}} \
\textbf{Qingkai Min}\textsuperscript{\rm{6}}  \\ \textbf{Zizhao Zhang}\textsuperscript{\rm{7}} \ \textbf{Yiwen Wang}\textsuperscript{\rm{1}} \ \textbf{Wenjie Li}\textsuperscript{\rm{4}} \ \textbf{Pengfei Liu}\textsuperscript{\rm{1,2,8}}\footnotemark[2] \\
\textsuperscript{1}Shanghai Jiao Tong University \
\textsuperscript{2}Shanghai Artificial Intelligence Laboratory \
\textsuperscript{3}Fudan University \\
\textsuperscript{4}The Hong Kong Polytechnic University \
\textsuperscript{5}Hong Kong University of Science and Technology \\
\textsuperscript{6}Westlake University \
\textsuperscript{7}Tsinghua University \
\textsuperscript{8}Generative AI Research Lab (GAIR) \\ 
\texttt{catchiz.1@sjtu.edu.cn, pengfei@sjtu.edu.cn}
}

\begin{document}
\maketitle
\begin{abstract}
The rapid growth of scientific literature imposes significant challenges for researchers endeavoring to stay updated with the latest advancements in their fields and delve into new areas. We introduce OpenResearcher, an innovative platform that leverages Artificial Intelligence (AI) techniques to accelerate the research process by answering diverse questions from researchers. OpenResearcher is built based on Retrieval-Augmented Generation (RAG) to integrate Large Language Models (LLMs) with up-to-date, domain-specific knowledge. Moreover, we develop various tools for OpenResearcher to understand researchers' queries, search from the scientific literature, filter retrieved information, provide accurate and comprehensive answers, and self-refine these answers. OpenResearcher can flexibly use these tools to balance efficiency and effectiveness. As a result, OpenResearcher enables researchers to save time and increase their potential to discover new insights and drive scientific breakthroughs. Demo, video, and code are available at: \url{https://github.com/GAIR-NLP/OpenResearcher}.
\end{abstract}

\begingroup
\renewcommand{\thefootnote}{\fnsymbol{footnote}}

\footnotetext[1]{Equal contribution.}
\footnotetext[2]{Corresponding author.}
\endgroup

\section{Introduction}
Global scientific publications are growing annually by about 4\%-5\% \citep{pinedo2024arzigo}, leading researchers to invest significant time and effort in thoroughly reviewing countless academic papers to find the knowledge that propels their research. This involves daily engagement with a wide range of literature to stay updated with the latest developments in their field, which is essential for maintaining the relevance and innovation of their work.

Recognizing the challenges and inefficiencies inherent in this process, considerable academic efforts have focused on AI-assisted scientific research \citep{wang2023scientific,zhai2023chatgpt}. They aim to answer the researcher questions from both junior and senior researchers. These questions can be broadly classified into three categories: (1) Scientific Question Answering \citep{pappas-etal-2020-biomrc,ruggeri-etal-2023-dataset,lee2023qasa,pramanick2024spiqa}, which seeks detailed information or clarification within a specific domain; (2) Scientific Text Summarization \citep{wang-etal-2022-multi,ding-etal-2023-cocoscisum,takeshita-etal-2024-aclsum,hsu2024chime,zhang2024masswnewdatasetbenchmark}, aimed at condensing the latest findings and developments into comprehensive overviews; and (3) Scientific Paper Recommendation \citep{bai2019scientific,kreutz2022scientificpaperrecommendationsystems,stergiopoulos2024academic,pinedo2024arzigo}, which involves suggesting relevant literature and studies based on the researcher's interests or current inquiries. However, academic applications typically focus on a \textbf{single} task, lacking a unified solution for all questions, allowing researchers to pose any inquiry freely.

Conversely, recent industry applications, like Perplexity AI,\footnote{\url{https://www.perplexity.ai/}} iAsk,\footnote{\url{https://iask.ai/}} You.com,\footnote{\url{https://you.com/}} phind,\footnote{\url{https://www.phind.com/}} and SearchGPT,\footnote{\url{https://chatgpt.com/search}} allow users to inquire about anything beyond specific tasks. They use Retrieval-Augmented Generation (RAG) \citep{lewis2020retrieval} technique to offer an innovative integration of generative Large Language Model (LLM) with web search capability. The core idea behind them is to offer users not just any answer, but the most accurate and contextually relevant information available. However, the \textbf{proprietary} nature of industry applications has hindered their development and may impede academic research in this field.

\begin{figure*}[t]
    \centering
    \includegraphics[width=0.8\linewidth]{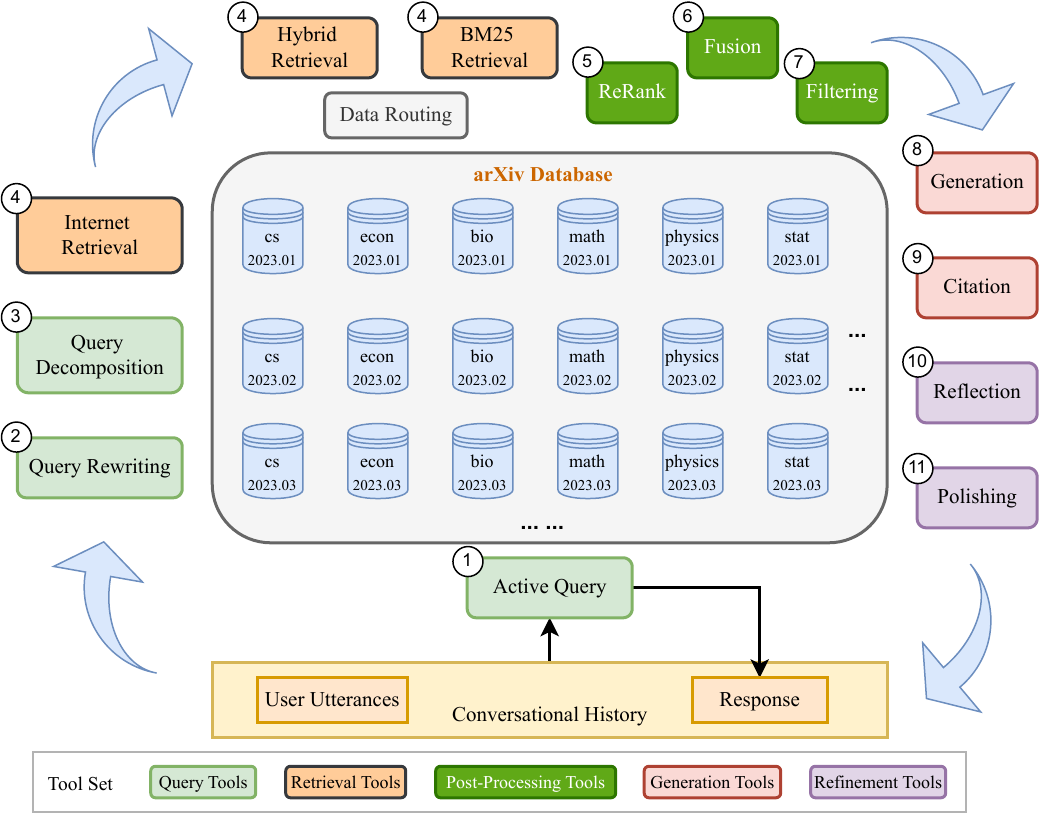}
    \caption{Main Workflow of OpenResearcher. }
    \label{fig:workflow}
\end{figure*}

Besides, both academic and industry applications serve as \textbf{passive} assistants, focusing solely on responding to user inquiries rather than engaging in active communication. To address these issues in academic and industry contexts, we developed OpenResearcher, an open-source project that harnesses AI to accelerate scientific research. Its main workflow is shown in Figure \ref{fig:workflow}, and its main contributions are as follows:
\begin{itemize}
\item \textbf{Unified Application} OpenResearcher can address researchers' diverse questions, such as Scientific Text Summarization, Scientific Paper Recommendation, etc.

\item \textbf{Open-Source} OpenResearcher is an impressive open-source system to rival the performance of industry applications.

\item \textbf{Active Assistant} OpenResearcher can connect in the mind or imagination to pose heuristic questions, guiding users to clarify queries for capturing their intent. 

\item \textbf{Retrieval Augmented} OpenResearcher can retrieve from the Internet and arXiv corpus to provide up-to-date, domain-specific, verified knowledge as supporting evidence. 

\item \textbf{Fexible Tool Usage} OpenResearcher can flexibly utilize bespoke tools to build a workflow for a better answer. For example, OpenResearcher adaptively calls a refinement tool to refine its initial outcomes. This approach helps avoid the computational cost associated with the unnecessary use of some tools.

\item \textbf{Conversational Interaction} OpenResearcher enables users to engage in deep discussions through conversational follow-up questions.

\end{itemize}

\section{Related Work}

\subsection{Academic Works}
Academic works for scientific research target a specific task, including Scientific Question Answering, Scientific Text Summarization, and Scientific Paper Recommendation. 

\noindent{\textbf{Scientific Question Answering}} generates answers for questions within extensive scientific articles. In the early days, cloze-style paper question answering datasets, such as emrQA \citep{pampari-etal-2018-emrqa}, BioRead \citep{pappas-etal-2018-bioread} and BioMRC \citep{pappas-etal-2020-biomrc}, are automatically created with the pre-defined question formats \citep{kwiatkowski-etal-2019-natural}. On the other hand, PubMedQA \citep{jin-etal-2019-pubmedqa}, BioAsq \citep{krallinger2020bioasq} and QASPER \citep{dasigi-etal-2021-dataset} involve human annotators in question creation. However, the questions are based only on abstracts. Recently, QASA \citep{lee2023qasa} offers advanced questions with annotators reading the entire paper. KIWI \citep{xu2024kiwi} uses expert and LLM interactions to refine initial answers into improved long-form answers. SPIQA \citep{pramanick2024spiqa} expands text question answering to multimodal question answering.

\noindent{\textbf{Scientific Text Summarization}} aims to condense the long scientific articles into a concise summary. Early works primarily focus on a knowledge graph-centric view \citep{wang-etal-2022-multi}. Recently, \citet{ding-etal-2023-cocoscisum} present CocoSciSum, a novel toolkit for controlled summarization of scientific documents, tailored to the scientific community's needs. \citet{takeshita-etal-2024-aclsum} introduce ACLSum, an expert-curated dataset for multi-aspect summarization of scientific papers, thoroughly covering challenges, approaches, and outcomes. \citet{hsu2024chime} release CHIME, a dataset that hierarchically organizes scientific studies to facilitate the generation of literature reviews. \citet{zhang2024masswnewdatasetbenchmark} introduces MASSW, a comprehensive dataset for summarizing multi-aspects of scientific workflows.

\noindent{\textbf{Scientific Paper Recommendation}} assists researchers in discovering relevant and suitable scientific information through recommendations. Early approaches \citep{tanner2019paper,ma2019personalized,sakib2020collaborative,manju2020cold} in Big Scholarly Data \citep{khan2017survey} have evolved into recently proposed hybrid recommender systems. \citet{pinedo2024arzigo} develop ArZiGo, a web-based prototype system for searching, managing, and recommending scientific articles. \citet{stergiopoulos2024academic} present a novel multi-stage recommendation system employing clustering, graph modeling, and deep learning, capable of operating on a large-scale scientific digital library with millions of users and papers.

However, these academic efforts focus on a single function without a unified solution for diverse inquiries and lack a user-friendly web application.

\subsection{Industry Research Applications}
Recent advancements in LLMs have prompted the industry to explore AI assistants for scientific research, like Perplexity AI, iAsk, You.com, phind, and SearchGPT, designed to handle all kinds of research inquiries in a dialogue. These applications combine chatbot-driven search engines with LLMs, which is academically termed Retrieval Augmented Generation (RAG). These applications also provide citations for the evidence behind their responses. However, the closed-source nature has limited their development and academic research in this area. 

\section{OpenResearcher}
OpenResearcher is designed to leverage AI to speed up the research process by efficiently responding to researchers’ inquiries. As shown in Figure \ref{fig:workflow},  OpenResearcher employs RAG to combine LLMs' internal knowledge with the latest external information. We design a Data Routing strategy for quick and precise information retrieval that can meet time and domain requirements. Lastly, we have developed multiple tools, including query tools, retrieval tools, post-processing tools, generation tools, and refinement tools. OpenResearcher can flexibly use these tools to customize a workflow for each query.

\subsection{Query Tools}
A key challenge in retrieval is its dependence on the user's initial query, which, if imprecise or vague, leads to ineffective results. Junior researchers may struggle to articulate their questions, and scientific terms used across different disciplines add to this complexity. To address this, we have developed tools to help define straightforward questions. 

\noindent\textbf{Active Query} OpenResearcher enhances a query by adding extra content and context. It asks users to specify their interest area or discipline. It can ensure that generated answers are highly relevant by covering nuances not initially mentioned.

\noindent\textbf{Query Rewriting} The users' queries are always suboptimal for retrieval, especially in real-world scenarios. Besides, the queries are commonly entailed in complex conversational interactions. Therefore, OpenResearcher rewrites the queries for better clarity and effectiveness.

\noindent\textbf{Query Decomposition} OpenResearcher decomposes the complex query into a series of sub-queries, improving precision and efficiency for more satisfying responses. Then each sub-query is processed by information retrieval and LLM generation systems accordingly to get the sub-answer.

\subsection{Retrieval Tools}
OpenResearcher uses advanced retrieval tools to gather comprehensive and accurate information from the Internet and arXiv corpus.

\noindent\textbf{Internet Retrieval} OpenResearcher conducts Internet Retrieval through search engines API to collect relevant online information.

\noindent\textbf{Hybrid Retrieval} OpenResearcher supports Hybrid Retrieval that employs sparse vector and dense vector representations of both queries and documents. By leveraging these compact vector embeddings, Hybrid Retrieval can more effectively capture semantic similarities and improve the relevance of retrieved documents. 

\noindent\textbf{BM25 Retrieval} OpenResearcher conducts BM25 Retrieval, an advanced algorithm used by search engines to rank documents based on their relevance to a query, factoring in term frequency and document length. BM25 stands out for its effectiveness in handling various search queries, making it a widely adopted method in information retrieval. 

\subsection{Data Routing Strategy}
We develop an advanced Data Routing strategy aimed at optimizing the performance of our hybrid retrieval tool. This retrieval tool currently requires substantial processing times to calculate the similarity between a query and all arXiv paper chunks, which can be resource-intensive. 

To address this issue, our strategy is to stratify the data based on both temporal and domain-specific information found in the metadata of the arXiv papers. It distributes data across multiple specialized databases, each aligned with a particular time frame and domain. Consequently, the retrieval tool only scans databases relevant to the query, which speeds up the search process and improves result accuracy by concentrating on the applicable data sets.

\subsection{Post-Processing Tools}
We develop Post-Processing Tools to rerank, fuse, and filter retrieved information,  removing noise and redundancy to provide the most pertinent outcomes for the generation of LLMs.

\noindent\textbf{Reranking}: OpenResearcher can use a reranking tool to reorder document chunks, prioritizing the most relevant results to condense the retrieval pool. 

\noindent\textbf{Fusion}: OpenResearcher can use a fusion tool to fuse the retrieved content from the same source into a single paragraph to enhance the context.

\noindent\textbf{Filtering}: OpenResearcher can use a filtering tool to filter out redundant and noisy content to preserve the most relevant information.

\subsection{Generation Tools}
OpenResearcher uses advanced LLMs to produce responses using retrieved information.

\noindent\textbf{Generation}
OpenResearcher prompts LLMs to utilize retrieved information to generate appropriate responses for user queries.

\noindent\textbf{Citation}
OpenResearcher can use a citation tool that employs the BM25 matching algorithm to link retrieved information with the response sentences, providing citations for each.

\subsection{Refinement Tools}
OpenResearcher utilizes LLMs to reflect and polish the initial responses, guaranteeing their accuracy and completeness.

\noindent\textbf{Reflection}
OpenResearcher prompts LLMs to evaluate the accuracy and completeness of generated responses, meanwhile highlighting grammatical and semantic flaws.

\noindent\textbf{Polishing}
OpenResearcher instructs LLMs to polish responses according to feedback received.

\section{Demonstration}

Our web application is built with Streamlit.\footnote{\url{https://streamlit.io/}} Our databases encompass arXiv publications from Jan. 2023 to Jun. 2024, enriched with metadata. This is because most LLMs are trained on pre-2023 data, enabling them to retain this information. This fact also inspires OpenResearcher to answer simple questions without any retrieval, only using LLMs' internal knowledge. We utilize the state-of-the-art GTE-large model \citep{li2023towards} as dense vector and efficient-splade-VI-BT-large \citep{10.1145/3477495.3531833} as sparse vector to vectorize our queries and paper chunks. These vectors serve for Hybrid Retrieval, and we use Qdrant\footnote{\url{https://qdrant.tech/}} for the vector storage. This Hybrid Retrieval tool extracts the 30 most similar chunks from each selected database. Elasticsearch\footnote{\url{https://github.com/elastic/elasticsearch}} supports our implementation of BM25 retriever, which extracts up to 80 chunks. The Bing\footnote{\url{https://www.bing.com/}} API finds 10 relevant outcomes for the Internet Retrieval tool. Besides, we utilize bge-reranker-v2-m3\footnote{\url{https://huggingface.co/BAAI/bge-reranker-large}} to implement our Reranking tool. This Reranking tool reduces the number of retrieved chunks to 10. Lastly, we use DeepSeek-V2-Chat \citep{deepseekai2024deepseekv2strongeconomicalefficient} as our backbone LLM to implement all LLM-powered tools, while also supporting various online LLM APIs and locally deployed LLMs through Ollama.\footnote{\url{https://ollama.com/}}

\begin{figure}[!ht]
    \centering
    \includegraphics[width=\linewidth]{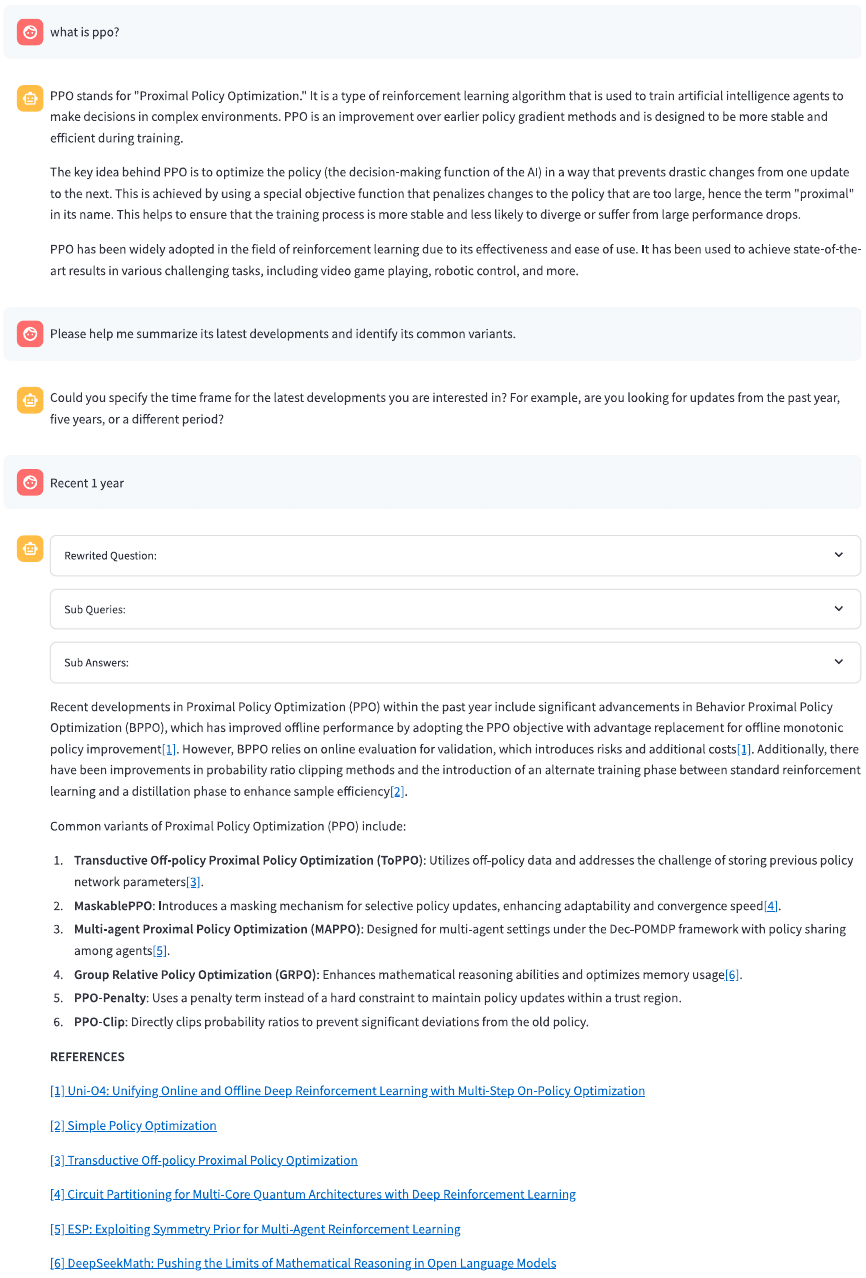}
    \caption{Case between user and OpenResearcher.}
    \label{fig:case_}
\end{figure}

Figure \ref{fig:case_}, whose completed screenshot is shown in Figure \ref{fig:c_case_} of Section \ref{sec:appendix}, demonstrates the strong capability of OpenResearcher. Firstly, OpenResearcher can flexibly construct a tailored workflow for different queries, including simple queries and complex queries. For simple questions like ``What is PPO?'', it directly employs LLMs to produce answers. For more complex queries like ``Summarize the recent latest developments and variants of PPO?'', it utilizes multiple tools and provides users with essential details, including active queries, rewritten query, decomposed sub-queries and their sub-answers, retrieved outcomes of each sub-query after post-processing, generated final answer, and citation. This example can showcase its flexibility in handling different queries. With this benefit, our OpenResearcher can speed up responses and reduce computational costs.

Secondly, this figure also shows OpenResearcher can pose questions to users for query clarification. Different from previous passive applications that only answer questions, OpenResearcher utilizes LLMs' internal knowledge to help users specify their question details. This tool is very crucial for junior students who often struggle to clearly express their questions and confusion. 

Thirdly, Figure \ref{fig:case_} demonstrates that OpenResearcher supports conversational question answering, enabling users to engage in multi-turn dialogues. This feature allows for continuous and deeper discussions within OpenResearcher.

Lastly, this figure shows our OpenResearcher can enhance the quality and reliability of generated content by retrieving supporting evidence from the Internet and arXiv corpus. Additionally, we have developed a citation tool that links the generated text to the retrieved information, making it easy for researchers to verify the sources and delve deeper by reading the original papers. 

\section{Experiment}
\subsection{Evaluation Data}
We have collected 109 research questions from more than 20 graduate students, comprising 38 questions on scientific paper recommendation, 38 on scientific text summarization, and 33 on others. These questions arise in their daily scientific research across areas including multimodal, agent, LLM alignment, tool learning, LLM safety, RAG, and others. Answers to these questions are commonly complex and lengthy, requiring graduate students to review many papers. Due to the considerable effort and cost of annotating ground truth answers, we opt to conduct a pairwise comparison instead of providing annotated ground truths. 

\subsection{Evaluation Applications}
Our baseline includes recent industry applications, containing Perplexity AI, iAsk, You.com, and Phind, complemented by a Naive RAG that only utilizes our hybrid retrieval and LLM generation tools. Regarding our OpenResearcher, we remove the Active Query tool to directly obtain the answer. Our OpenResearcher flexibly uses these tools to generate answers without the need to follow the main workflow sequentially.

\begin{table*}[!ht]
\centering
\small
\setlength{\tabcolsep}{10pt}
\begin{tabular}{lcccccccccc}
\toprule
   \multirow{2}{*}{\textbf{Models}}       & \multicolumn{3}{c}{\textbf{Correctness}} & \multicolumn{3}{c}{\textbf{Richness}} & \multicolumn{3}{c}{\textbf{Relevance}} \\ \cmidrule(lr){2-4} \cmidrule(lr){5-7} \cmidrule(lr){8-10}
          & \textbf{Win}          & \textbf{Tie}  & \textbf{Lose}       & \textbf{Win}          & \textbf{Tie}  & \textbf{Lose}   & \textbf{Win}          & \textbf{Tie}  & \textbf{Lose}  \\ \midrule
\texttt{Ask} & 2         & 16        & 12       & 12  & 6        & 12   & 2 & 8 & 20  \\
\texttt{You.com} & 3         & 21        & 6     & 9  & 5        & 16   & 4 & 13 & 13   \\ 
\texttt{Phind} & 2         & 26        & 2     & 15  & 7        & 8   & 5 & 13 & 12   \\ \midrule
\texttt{Naive RAG} & 1        & 22        & 7       & 14  & 8       & 8  & 5 & 16 & 9  \\
\texttt{OpenResearcher} & \textbf{10}   & 13   & 7  & \textbf{25}  & 4        & 1   & \textbf{15} & 13 & 2 \\
\bottomrule    
\end{tabular}
  \caption{Human Preference compared with Perplexity AI outcome. ``Win'' means that the current method beats Perplexity AI. More ``Win'' times means a superior application.}
  \label{tab:human_benchmark}
\end{table*}

\subsection{Evaluation Metric}
In all evaluations, we compared the candidate outcomes from Naive RAG, OpenResearcher, iAsk, You.com, and Phind with those from Perplexity AI. If the candidate outcome outperforms Perplexity AI, it is notated as a ``Win''. 

We evaluate the generations from the three quality dimensions:
(1) \textbf{Information Correctness} assesses the factual accuracy of the answers provided by the candidates. It is critical to determine if the information in each output is correct, as inaccuracies can severely undermine the utility of a QA system. (2) \textbf{Information Richness} involves evaluating the depth and scope of the information provided in the answers. Information richness captures whether an answer provides a thorough explanation or context beyond just addressing the question directly. (3) \textbf{Information Relevance} judges whether the information presented in the outputs is directly relevant to the question asked. Even if an answer is rich in information and correct, it may not be useful if it does not directly address the query. 

\subsection{Human Preference}
We engaged 12 students with good research experience to conduct the human evaluation. Given the complexity of research questions, we randomly selected 30 questions for human evaluation, ensuring equal coverage of scientific question answering, scientific text summarization, and scientific paper recommendation. For quality control, each instance is annotated by two annotators whose agreement is measured. A third annotator can be involved to resolve disagreements between the two annotators.

The result is shown in Table \ref{tab:human_benchmark} with an overall agreement of 90.67\%. Our OpenResearcher achieves superior information correctness, relevance, and richness compared to all other applications. OpenResearcher significantly outperforms Perplexity AI with more ``Win'' than ``Lose''. Specifically, compared to Naive RAG, OpenResearcher demonstrates better performance in all metrics. This suggests that our various tools significantly enhance the quality of the answers. 

\subsection{LLM Preference}
\begin{table}[!ht]
\centering
\small
\setlength{\tabcolsep}{4pt}
\begin{tabular}{lccccccc}
\toprule
   \multirow{2}{*}{\textbf{Models}}   & \multicolumn{3}{c}{\textbf{Richness}} & \multicolumn{3}{c}{\textbf{Relevance}} \\ \cmidrule(lr){2-4} \cmidrule(lr){5-7} 
               & \textbf{Win}          & \textbf{Tie}  & \textbf{Lose}   & \textbf{Win}          & \textbf{Tie}  & \textbf{Lose}  \\ \midrule
\texttt{iAsk}   &  42  & 0        & 67   & 38 & 0 & 71  \\
\texttt{You.com}   & 15  & 0        & 94   & 16 & 0 & 93   \\ 
\texttt{Phind}   &  52  & 1       & 56   & 54 & 0 & 55   \\ \midrule
\texttt{Naive RAG}    & 41  & 1       & 67  & 57 & 0 & 52  \\
\texttt{OpenResearcher}  & \textbf{62}  & 2        & 45   & \textbf{74} & 0 & 35 \\
\bottomrule    
\end{tabular}
  \caption{GPT-4o Preference Results compared with Perplexity AI outcome. }
  \label{tab:gpt_benchmark}
\end{table}
Inspired by the widespread use of GPT-4 series for pairwise comparison \citep{zheng2023judging,wang2023pandalm,sun2024promptchainingstepwiseprompt} and their different preferences compared to humans \citep{li2024dissecting}, we also utilize GPT-4o for LLM preference evaluation. We evaluate based on two criteria: information richness and relevance, since GPT-4o struggles to verify information accuracy without external knowledge. Despite the availability of citation papers, their length and quantity exceed LLMs' capacity to confirm factuality.

The results are shown in Table \ref{tab:gpt_benchmark}. This supplemental LLM evaluation further demonstrates our system's powerful performance. These results show our OpenResearcher achieves the best information relevance and richness among all applications. Furthermore, OpenResearcher surpasses Naive RAG in both metrics, demonstrating its superior performance due to our design. 

\section{Conclusion}
We introduce OpenResearcher, an active AI assistant to accelerate the research process, catering to a broad spectrum of inquiries from researchers. OpenResearcher employs Retrieval-Augmented Generation (RAG) to enhance LLMs with the latest, verified, and domain-specific knowledge. It interacts with users to clarify their queries. Moreover, we have developed various tools for OpenResearcher to understand researchers' queries, search from the scientific literature, filter retrieved information, provide accurate and comprehensive answers, and refine these answers. OpenResearcher can use these tools flexibly to build a pipeline that delivers accurate and comprehensive answers, outperforming those from industry applications, as judged by human and GPT-4o.

\section*{Ethical Considerations}
OpenResearcher integrates LLMs and search engines, known as retrieval-augmented generation (RAG), to accelerate scientific research. Despite being instructed to ground the generated responses in retrieved knowledge from scientific publications, LLMs may still generate hallucinations. Consequently, users are advised to verify crucial information derived from our LLM-based features.

\section*{Acknowledgments}
The authors would like to thank the anonymous reviewers for their suggestions and feedback on the work. This work was partially funded by the National Natural Science Foundation of China (62476168), Qingyuan Research Project.

\bibliography{anthology,custom}

\appendix
\section{Completed Case}
\label{sec:appendix}
\begin{figure*}[!ht]
    \centering
\includegraphics[width=0.65\linewidth]{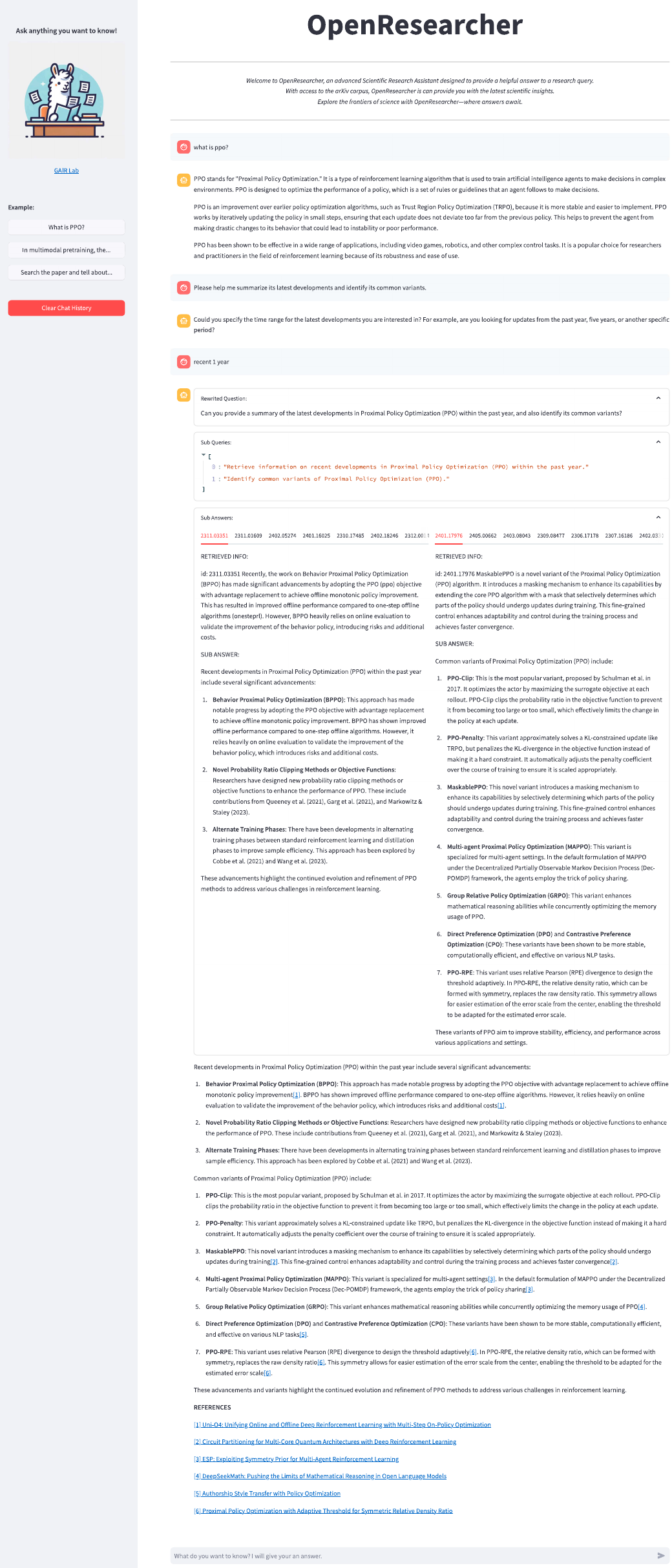}
    \caption{Screenshot showing the completed case in Figure \ref{fig:case_}.}
    \label{fig:c_case_}
\end{figure*}

\end{document}